\documentclass[onecolumn,prd,superscriptaddress,showpacs,floatfix,%  
preprintnumbers,nofootinbib,showkeys]{revtex4}  

\usepackage{amsmath,amsfonts}
\usepackage{graphicx}
\def\issue(#1,#2,#3){#1, #2 (#3)} % AIP format

\def\APP(#1,#2,#3){{\rm Acta Phys.\ Polon.} \ \issue({\bf #1},#2,#3)}
\def\ANP(#1,#2,#3){{\rm Annals of Physics} \ \issue({\bf #1},#2,#3)}
\def\ARNPS(#1,#2,#3){{\rm Ann.\ Rev.\ Nucl.\ Part.\ Sci.} \ \issue({\bf #1},#2,#3)}
\def\CPC(#1,#2,#3){{\rm Comp.\ Phys.\ Comm.} \ \issue({\bf #1},#2,#3)}
\def\CIP(#1,#2,#3){{\rm Comput.\ Phys.} \ \issue({\bf #1},#2,#3)}
\def\EPJ(#1,#2,#3){{\rm Eur.\ Phys.\ J.} \ \issue({\bf #1},#2,#3)}
\def\EPJD(#1,#2,#3){Eur.\ Phys.\ J. Direct\ C \ \issue({\bf #1},#2,#3)}
\def\IJMP(#1,#2,#3){{\rm Int.\ J.\ Mod.\ Phys.} \ \issue({\bf #1},#2,#3)}
\def\JHEP(#1,#2,#3){{\rm J.\ High Energy Physics} \ \issue({\bf #1},#2,#3)}
\def\JP(#1,#2,#3){{ J.\ Phys.} \ \issue({\bf #1},#2,#3)}
\def\MPL(#1,#2,#3){{Mod.\ Phys.\ Lett.} \ \issue({\bf #1},#2,#3)}
\def\NP(#1,#2,#3){{Nucl.\ Phys.} \ \issue({\bf #1},#2,#3)}
\def\NIM(#1,#2,#3){{ Nucl.\ Instrum.\ Meth.} \ \issue({\bf #1},#2,#3)}
\def\PL(#1,#2,#3){{ Phys.\ Lett.} \ \issue({\bf #1},#2,#3)}
\def\PR(#1,#2,#3){{ Phys.\ Rev.} \ \issue({\bf #1},#2,#3)}
\def\PRL(#1,#2,#3){{ Phys.\ Rev.\ Lett.} \ \issue({\bf #1},#2,#3)}
\def\SJNP(#1,#2,#3){{ Sov.\ J. Nucl.\ Phys.} \ \issue({\bf #1},#2,#3)}
\def\ZP(#1,#2,#3){{Zeit.\ Phys.} \ \issue({\bf #1},#2,#3)}
\def\be {\begin{equation}}
\def\ee {\end{equation}}
\def\bea {\begin{eqnarray}}
\def\eea {\end{eqnarray}}
\def\bsbsbar {B_s^0-{\bar{B}}_s^0}

\def\dzbar {\bar{D}{}^0}
\def\bs {B_s^0}
\def\bsbar {{\bar{B}}_s^0}
\def\bra {\langle}
\def\ket {\rangle}
\def\dgs {\Delta\Gamma_s}
\def\dms {\Delta M_s}

\begin{document}
%\begin{titlepage}
%\tableofcontents
%\pagestyle{empty}
%\begin{center}
%{\Large {\bf
\title{$B_s\to D_s K$ as a Probe of CPT Violation}% \\[5mm]
\bigskip

\author{Anirban Kundu}
\affiliation{Department of Physics, University of Calcutta, \\
92, Acharya Prafulla Chandra Road, Kolkata 700009, India}
\author{Soumitra Nandi}
\affiliation{Theoretische Elementarteilchenphysik, Naturwissenschaftlich Technische Fakult\"at,\\
Universit\"at Siegen, 57068 Siegen, Germany}
\author{Sunando Kumar Patra}
\affiliation{Department of Physics, University of Calcutta, \\
92, Acharya Prafulla Chandra Road, Kolkata 700009, India}
\author{Amarjit Soni}
\affiliation{Physics Department, Theory Group, Brookhaven National Laboratory, Upton, NY 11973, USA}

\begin{abstract}

We discuss some possible signals of CPT violation in the $B_s$ system 
that may be probed at the Large Hadron Collider (LHC).
We show how one can construct combinations of
observables coming from tagged and untagged decay rates of $B_s\to D_s^\pm K^\mp$
that can unambiguously differentiate
between CPT violating and CPT conserving new physics (NP) models contributing in $\bsbsbar$
mixing. We choose this particular mode as an illustrative example for two reasons: (i) In the 
Standard Model, there is only one decay amplitude, so it is easier to untangle any new physics; 
(ii) $B_s$ being a neutral meson, it is possible to unambiguously identify any sign of CPT violation 
that occurs only in mixing but not in decay. We define an observable which is useful to extract the CPT violating parameter in 
$B_s$ decay, and also discuss how far the results are applicable even if CPT violation is 
present in both mixing and decay.

\end{abstract}

\date{\today}
\maketitle

%%%%%%%%%%%%%%%%%%%%%%%%%%%%%%%%%%%%%%%%%%%%%%%%%%%%%%%%%%%%%%%%%%%%%%%%%%%%%%

\section{Introduction}
\label{sec1}

%%%%%%%%%%%%%%%%%%%%%%%%%%%%%%%%%%%%%%%%%%%%%%%%%%%%%%%%%%%%%%%%%%%%%%%%%%%%%%

The combined discrete symmetry CPT, taken in any order, is an exact symmetry of any axiomatic quantum field theory (QFT).
%both in the Hamiltonian quantum field theory (QFT) and in any local axiomatic QFT. 
CPT conservation is indeed supported by the experiments;
all tests for CPT violation (CPTV) that have been done so far \cite{pdg} have yielded null results,
consistent with no CPTV, and very stringent limits on CPTV parameters have been obtained
\cite{kostel001} in different systems. The only possible exception is the apparent mass difference
between the top quark and its antiparticle as obtained by the CDF collaboration in Fermilab \cite{cdf-cpt}:
\be
m_t - m_{\bar t} = -3.3 \pm 1.7~{\rm GeV}\,,
\ee
but other experiments got results which are consistent with zero, and so is the world average \cite{wa-cpt}:
\be
m_t - m_{\bar t} = [-0.44 \pm 0.46 ~({\rm stat.}) \pm 0.27~({\rm syst.})]~{\rm GeV}\,.
\ee
What, then, should be the motivation to investigate the possibility of
CPTV particularly in the B system? There are three main reasons:
\begin{itemize}
\item
Any symmetry which is supposed to be exact ought to be questioned.
We may get a surprise, just like the discovery of CP violation. CPTV may
very well be flavor-sensitive, and so the constraints obtained from the
K system \cite{nussinov} may not be applicable to the B systems. There is still the 
possibility of a sizable CPTV in the B systems. If there is some tension between the data and
the Standard Model (SM)
expectations, we should ask whether this is due to CPTV, or a more canonical CPT conserving
new physics (NP).

\item
For the bound systems like mesons, asymptotic states, whose existence is a
prerequisite for the CPT theorem, are not
uniquely defined \cite{quark}. Quarks and gluons are bound inside the hadrons and cannot be
considered, in a true sense, asymptotic states. 

\item
Some nonlocal and nonrenormalizable string-theoretic effects 
%\hldc{(For AS: Warped ED models need not necessarily lead to this. 
%In fact, they need not have anything to do with string theories.)} 
may appear
at the Planck scale with a possible ramification at
the weak scale through the effective Hamiltonian \cite{cpt-hamil}. CPTV through such non-local
interacting QFT does not necessarily lead to the violation of Lorentz symmetry \cite{novikov}.

\end{itemize}

Recently the issue of CPTV has received more attention due to the growing phenomenological importance
of CPT violating scenarios in neutrino physics and in cosmology \cite{others}.
It is also necessary to find some observables that will clearly discriminate CPT violating signals 
from CPT conserving ones. 
A comprehensive study of CPTV in the neutral K meson system, with a formulation
that is closely analogous to that in the B system, may be found in \cite{lavoura}.

CPTV in the B systems, and its possible signatures, have been already investigated
by several authors \cite{datta,patra}. It was shown that the
lifetime difference of
the two mass eigenstates, or the direct CP asymmetries and semileptonic observables,
may be affected by such new physics. The experimental limits are set by both BaBar, who looked
for diurnal variations of CP-violating observables \cite{cpt-babar}, and Belle, who looked for
lifetime difference of $B_d$ mass eigenstates \cite{cpt-belle}. This makes it worthwhile
to look for possible CPTV effects in the $B_s$ system (by $B_s$ we generically mean both
$\bs$ and $\bsbar$ mesons).

In this paper, we would like to investigate the signatures of CPT violation
in the $B_s$ system, both in $\bsbsbar$ mixing and in $B_s$ decays. We would like 
to emphasize that this 
is a model-independent approach in the sense that we do not specify any definite model 
that might lead to CPT violation; in fact, as far as we know, all studies on CPT violation are 
based on some phenomenological Lagrangian to start with.

As an illustrative
example, we consider the nonleptonic $\bs (\bsbar) \to D_s^+ K^-$ and $\bs (\bsbar) \to D_s^- K^+$ decays.
The $\bsbar$ decays are mediated by color-allowed tree-level transitions $b\to u\bar{c}s$ and
$b\to c\bar{u}s$. These are single-amplitude processes in the SM, so that any non-trivial contribution
beyond the SM expectations, like direct CP asymmetry, is a clear signal of NP.
This set of channels is also of interest as 
in the SM, both the amplitudes are of same order,
${\cal O}(\lambda^3)$
in the standard Wolfenstein parametrization of the CKM matrix (so that the event rates are comparable), 
and same final states can be reached both from $\bs$ and $\bsbar$. 
The importance of such modes to unveil any NP has
already been emphasized; {\em e.g.}, see \cite{fleischer-0304027,nandi1,nandi2,debruyn}.
The decay was first observed by the 
CDF and the Belle collaborations \cite{cdf-0809.0080,belle-0809.2526},
and recently the LHCb collaboration has measured the branching ratio to be
\cite{LHCb-1204.1237}
\be
{\rm Br}(B_s\to D_s^\mp K^\pm) = (1.90 \pm 0.23)\times 10^{-4}
\ee
where the errors have been added in quadrature.
We also note that flavor-specific NP in these
channels is relatively unconstrained \cite{bauer}.
LHCb has also measured several time-dependent CP violating observables in $B_s \to
D_s^\mp K^\pm$ using flavor-tagged and flavor-untagged observables \cite{LHCb-029}.

Here we do a more general analysis considering both the CPT violating and CPT conserving NP
contributions to $\bsbsbar$ mixing. We show how one can construct combinations of
observables coming from tagged and untagged decay rates that can unambiguously differentiate
between CPT violating and CPT conserving NP models.
On the other hand, if there is some CPTV contribution only to $B_s$ decays, it might be difficult 
to differentiate it from CPT conserving NP in this approach. 
%However, in the $B_s$ decay, we consider only CPTV and 
%discuss the impact of the presence of CPT conserving NP.
We define an observable which is useful to extract the CPT violating parameter in decay.

We will consider both these cases separately:
first, when CPTV (or CPT conserving NP)
is present only in the operators responsible for decay but not in those responsible for the mixing;
and second, when the same is present also in the $\bsbsbar$
mixing amplitude. As we will show explicitly, the extraction of CPTV in mixing is independent of
the CPTV in decay and any other CPT conserving NP either in decay or mixing.

 The first possibility of NP (including CPT violation) only in decay
can arise
if the NP operators are strongly flavor-dependent, like those in R-parity violating supersymmetry,
or leptoquark models. As we are considering final states that can be accessed both from $\bs$ and $\bsbar$,
any such NP will necessarily contribute in $\bsbsbar$
mixing, in particular to its absorptive part, and will change the decay width
difference $\dgs$. 
Apart from the short-distance contributions to the absorptive part, 
there can be non-negligible long-distance effects too, coming from mesonic intermediate states 
\cite{chua}.
However, the accuracy of the present data on $\dgs$,
the lifetime difference of two $B_s$ mass eigenstates, is relatively weak.
The most accurate result comes from the LHCb collaboration \cite{lhcb-delgamma}:
$\dgs/\Gamma_s = 0.176 \pm 0.028$.
Even the SM prediction \cite{lenz-bs} has a
large uncertainty. Thus, as a first approximation, one can consider such NP effects
only in decay and not in mixing, where it is in all probability subleading.

For the second case, one can construct several observables from the time-dependent
tagged and untagged decay rates, and some of them are identically zero if there is no CPTV
in mixing, irrespective of whether there is any CPTV  in decay, or some
CPT conserving NP.

%However, as we will show, the analysis goes through even if we have CPTV in both
%decay and mixing, as the observables extracted in one case is independent and unaffected
%by the observables extracted for the other case.  

The Belle Collaboration \cite{cpt-belle}
places limits on the CPTV parameters in mixing, but no such limits exist
for CPTV in decay. Also, the Belle limits are valid for the $B_d$ system, but one can
expect similar numbers for the $B_s$ system too, even if CPTV is flavor-dependent.
Like the experimental tests on CP-violation, various independent cross-checks
on CPTV are also essential.
Needless to say, one can play the same game with decays like
$B_s\to D^0\phi$ and $B_s\to \dzbar\phi$, and can form more observables (although not
independent of the original ones) out of the CP-eigenstates of $D^0$ and $\dzbar$
in the final state.

The paper is arranged as follows. In the next section, we outline the necessary
formalism for CPTV in decay, vis-a-vis that for the SM as well as CPT conserving NP.
We also construct observables that may indicate the presence of CPT violation (or
any NP in general). In Section
III, we do the same for CPT violation in $\bsbsbar$ mixing, including the construction
of observables that can differentiate CPTV and CPT conserving NP.
In Section IV, we summarize and conclude.

%%%%%%%%%%%%%%%%%%%%%%%%%%%%%%%%%%%%
\section{CPT Violation in Decay}
%%%%%%%%%%%%%%%%%%%%%%%%%%%%%%%%%%%%

\subsection{$\bsbsbar$ mixing and $B_s\to D_s^\pm K^\mp$ in the SM}

The $\bsbsbar$ mixing is controlled by the off-diagonal term $H_{12} = M_{12}
- (i/2)\Gamma_{12}$ of the $2\times 2$ Hamiltonian matrix, with the mass difference
between two mass eigenstates $B_H$ and $B_L$ given by (in the limit
$|\Gamma_{12}| \ll |M_{12}|$)
\be
\dms \equiv M_{sH}-M_{sL} \approx 2\vert M_{12}\vert\,,
\ee
and the width difference by
\be
\dgs \equiv \Gamma_{sL} - \Gamma_{sH} \approx 2\vert \Gamma
_{12} \vert\cos\phi_s\,,
\ee
where $\phi_s \equiv \arg(-M_{12}/\Gamma_{12})$. CPT conservation ensures
$H_{11}=H_{22}$.

The eigenstates are defined as
\be
|B_{H(L)}\ket = p|\bs\ket + (-) q|\bsbar\ket\,,
\label{b-eigen}
\ee
where $|p|^2+|q|^2=1$ is the normalization, and one defines
\be
\alpha\equiv q/p = \exp(-2\beta_s)
\ee
where $2\beta_s$ is the mixing phase of the $\bsbsbar$ box diagram.

For the single-amplitude decays $B_s\to D_s^\pm K^\mp$, the amplitudes are of the form
\bea
&& A(\bs\to D_s^+ K^-)  =  T_1 e^{i \gamma}\,,\ \ \ \ \ \  %\left(1 - y_f \right), \hskip 30pt
A(\bs \to  D_s^- K^+)  = T_2 \,, \nonumber \\
&& A(\bsbar \to D_s^+ K^-) =  T_2\,, \ \ \ \ \ \ \
A(\bar\bs \to D_s^- K^+)  = T_1 e^{- i \gamma}\,,
\label{sm-decay}
\eea
where $T_1$ and $T_2$ are real amplitudes times the strong phase, which we parametrize as
\be
\arg\left(\frac{T_1}{T_2}\right)  = \Delta\,,
%\left\vert\frac{T_2}{T_1}\right\vert\exp(i\Delta)\,,
\label{strphs}
\ee
and $\gamma = \arg(-V_{ud}V_{ub}^\ast/V_{cd}V_{cb}^\ast)$, so that to a very good approximation,
$V_{ub} \approx \vert V_{ub}\vert \exp(-i\gamma)$. The
quantity $\xi_f \equiv \alpha \bar{A}_f/A_f$, where $A_f \equiv  A(\bs\to D_s^+ K^-)$ and $\bar{A}_f \equiv
A(\bsbar \to D_s^+ K^-)$, carries a weak phase of $-(2\beta_s+\gamma)$.

Let us define, following \cite{fleischer-0304027},
\bea
\bra {\rm Br}(B_s\to D_s^+K^-)\ket &=& {\rm Br}(\bs\to D_s^+K^-) + {\rm Br}(\bsbar\to D_s^+K^-)\,,\nonumber\\
\bra {\rm Br}(B_s\to D_s^-K^+)\ket &=& {\rm Br}(\bs\to D_s^-K^+) + {\rm Br}(\bsbar\to D_s^-K^+)\,,
\eea
so that these untagged rates are the same in the SM, even though a future measurement of
the time-dependent branching fractions at the LHCb may show nonzero
CP violation.

%%%%%%%%%%%%%%%%%%%%%%%%%
\subsection{CPT violation in $B_s$ decay}
%%%%%%%%%%%%%%%%%%%%%%%%%
In order to take into account CPTV in decay, we parametrize various transition amplitudes for the decay
$B_s\to D_s^\pm K^\mp$ as \cite{dalitz,kooten}
\bea
&& A(\bs\to D_s^+ K^-)  =  T_1 e^{i \gamma} \left(1 - y_f \right)\,,\ \ \ \ \ \  %\left(1 - y_f \right), \hskip 30pt
A(\bs \to  D_s^- K^+)  = T_2 \left(1 + y^*_f \right)\,, \nonumber \\
&& A(\bsbar \to D_s^+ K^-) =  T_2  \left(1 - y_f \right)\,, \ \ \ \ \ \ \
A(\bar\bs \to D_s^- K^+)  = T_1 e^{- i \gamma}\left(1 + y^*_f \right)\,,
\label{cpt-decay}
\eea
where CPT violation (in decay) is parametrized by the complex parameter $y_f$, and $y_f$ is real if $T$ is conserved.
The CPT violation is proportional to the difference $A(\bs\to D_s^+ K^-)^{\ast} - A(\bar\bs \to D_s^- K^+)$ or
$A(\bsbar \to D_s^+ K^-)^{\ast} - A(\bs \to  D_s^- K^+)$.

We define the complete set of four relevant amplitudes, with $|f\ket\equiv |D_s^+K^-\ket$ and $|\bar f\ket
\equiv |D_s^-K^+\ket$,
\bea
&& \nonumber A_f = \bra f|H|\bs\ket\,,\ \ \
 A_{\bar f} = \bra \bar f|H|\bs\ket  \,,\nonumber\\
&&
\bar A_f = \bra f|H|\bsbar\ket \,,\ \ \
\bar A_{\bar f} = \bra \bar f|H|\bsbar\ket\,,
\eea
so that the ratios
\be
\xi_f = \alpha{\bar A_f}/{A_f}\,,\ \ \ \xi_{\bar f} = \alpha{\bar A_{\bar f}}/{A_{\bar f}}\,,
\ee 
are independent of $y_{f}$; the CPTV effect in the decays cancels in the ratio. 
%$\alpha$ parameterizes the $\bsbsbar$ mixing. 
We also have $|\xi_f| = 1 /|\xi_{\bar f}|$ and  $\arg(\xi_{f (\bar f)}) = -(2\beta_s+\gamma 
+(-) \Delta)$ where $\Delta$ is defined in eq.\ (\ref{strphs}).

From Eq.\ (\ref{cpt-decay}) we get
\bea
|A(\bs\to D_s^+ K^-)|^2 + |A(\bsbar \to D_s^+ K^-)|^2 &=& (|T_1|^2 + |T_2|^2)\,\left| 1-y_f \right|^2\,, \nonumber \\
|A(\bs\to D_s^- K^+)|^2 + |A(\bsbar \to D_s^- K^+)|^2 &=& (|T_1|^2 + |T_2|^2)\,\left| 1+y^*_f \right|^2\,.
\label{amp01}
\eea
Thus we can define an asymmetry
\be
A_{br}^{CPT}= \frac{\bra {\rm Br}(B_s \to D_s^+ K^-) \ket- \bra {\rm Br}(B_s \to D_s^- K^+)\ket}
{\bra {\rm Br}(B_s \to D_s^+ K^-)\ket + \bra {\rm Br}(B_s
 \to D_s^- K^+)\ket} = - 2\,\frac{{\rm Re}(y_f)}{1+|y_f|^2}\approx - 2 \, {\rm Re} (y_f)\,, ~{\rm for}~
|y_f|^2 \ll 1
\label{brfrac}
\ee
We have already seen that this asymmetry is zero in the SM. Using Eq.~(\ref{brfrac}), the real part of the
CPTV parameter $y_f$ can be directly probed from the difference of
the untagged rates (as the initial state $B_s$ flavor is summed over) 
${\rm Br}(B_s \to D_s^+ K^-)$ and ${\rm Br}(B_s \to D_s^- K^+)$.

One can have a rough idea of the LHCb reach in measuring ${\rm Re}(y_f)$. With 1 fb$^{-1}$ of integrated 
luminosity, LHCb has obtained $1390\pm 98$ events \cite{LHCb-029}. With full LHCb upgrade to an integrated 
luminosity of 50 fb$^{-1}$, total number of events should go up by a factor of about 200, as a twofold gain in 
the yield is expected when the LHC reaches $\sqrt{s}=13$-$14$ TeV (as the cross section of $pp \to
b\bar{b}X$ scales almost linearly with $\sqrt{s}$), and another twofold gain is expected in 
the trigger efficiency when the detector is upgraded. These $0.28$ million events should be roughly equally 
divided between $D_s^+K^-$ and $D_s^- K^+$. The advantage is that there is no need to tag the flavor of 
the initial $B_s$. The statistical fluctuation for each channel is about 375, 
and detection of CPT violation over such fluctuations results in a sensitivity of $375/140000 \approx 
0.0027$ for ${\rm Re}(y_f)$. Note that LHCb already has a plan to measure CPT violation in the decay 
$B^0\to J/\psi[\to \pi^\mp \mu^\pm \nu(\bar\nu)] K^0$ \cite{1208.3355}. 
However, in this estimate we have only concerned ourselves with the
statistical reach; we leave
it to the experimentalists to address the systematic errors.

%This is to be multiplied by the effective flavor 
%tagging efficiency of 1.9\%, which gives approximately 5300 flavor-tagged events. Thus, the 
%$1\sigma$ reach of ${\rm Re}(y_f)$ is $1/\sqrt{5300} = 0.013$ at the most.

Let us compare this to a case where there is no CPT violation, but some CPT conserving NP is present which
contributes to either $b\to u\bar{c}s$ or $b\to c\bar{u}s$ transitions, or maybe both. If this NP leads to
observable CP violating effects, we can write the 
various amplitudes for the $B_s\to D_s^\pm K^\mp$ decays as
\bea
&& A(\bs\to D_s^+ K^-)  =  T_1 e^{i \gamma} \left(1 + a~e^{i(\theta-\gamma+\sigma)} \right)\,,\ \ 
A(\bs \to  D_s^- K^+)  = T_2 \left(1 + a'~e^{i(\theta'+\sigma')} \right)\,,\nonumber\\
&&A(\bsbar \to D_s^+ K^-) =  T_2  \left(1 + a'~e^{-i(\theta'-\sigma')} \right)\,,\ \ 
A(\bar\bs \to D_s^- K^+)  = T_1 e^{- i \gamma}\left(1 +  a~e^{-i(\theta-\gamma-\sigma)} \right)\,.
\label{np-decay}
\eea
The amplitudes, obviously, are related by CP conjugation. The NP is parametrized by the (relative) amplitudes
$a$, $a'$, the new weak phases $\theta$, $\theta'$, and the new strong phase differences $\sigma$, $\sigma'$. 
Therefore, the asymmetry defined in Eq.~\ref{brfrac} is given by
\be
A_{br}^{NP} =  -2 \frac{\,a\,|T_1|^2 \sin(\theta-\gamma) \sin\sigma +a'\,|T_2|^2 \sin{\theta'}\sin{\sigma'}}
{|T_1|^2\left(1+a^2 + 2\,a\cos(\theta-\gamma) \cos\sigma\right) + |T_2|^2\left(1 + a'^2+2\,a'\cos{\theta'}\cos{\sigma'}\right)}.
\label{brfracnp}
\ee
Hence, a nonzero value of $A_{br}$ could be due to either CPTV or CPT conserving NP (which, perhaps, is flavor-dependent,
and definitely not of the minimal flavor violation type). 
As both the decays are color-allowed, one can even invoke the color-transparency argument 
\cite{bjorken} to claim that all strong phases
are small; but CPTV effects are not expected to be large either.

Eq.\ (\ref{brfrac}) is in general true for all decays which are either (i) single-amplitude in the SM, be it 
tree or penguin, or (ii) multi-amplitude in the SM but with one amplitude highly dominant over the others. 
Single-amplitude decays are preferred simply because any nonzero asymmetry as in Eqs.\~(\ref{brfrac}) 
or (\ref{brfracnp}) can be unambiguously correlated with NP. The same 
observable $A_{br}^{CPT}$ can be defined for charged B decays, or even D and K decays. However, in all cases, 
CPT conserving (but necessarily CP violating) NP can 
always mimic the asymmetry, unless there are strong motivations for the corresponding
amplitudes to be highly subdominant, or 
the strong phase difference between the two amplitudes to be zero or vanishingly small.

On the other hand, if there is CPT violation in mixing too, this formalism does not hold, because the definition of the 
mass eigenstates also contains CPT violating parameters (see later). In that case, we suggest using single-amplitude
charged B meson decay modes, like $B^+ \to D^0 K^+$ and $B^+ \to \overline{D}{}^0 K^+$.

If there is no other CPT conserving NP, but the $\bsbsbar$ mixing matrix has CPTV built in, the asymmetry 
is still nonzero, as the individual branching fractions are functions of the CPTV parameter $\delta$ (see below)
in the mixing matrix \cite{patra}.

%%%%%%%%%%%%%%%%%%%%%%%%%%%%%%%
\section{CPT Violation in Mixing}
%%%%%%%%%%%%%%%%%%%%%%%%%%%%%%%

This subsection closely follows the formulation developed in 
\cite{patra}, but let us quote some relevant expressions
for completeness. CPT violation in the Hamiltonian matrix is introduced through the complex parameter $\delta$:
\be
\delta = \frac{H_{22}-H_{11}}{\sqrt{H_{12}H_{21}}}\,, \label{deldef}
\ee
so that the Hamiltonian matrix looks like
\be
  \mathcal H = \left[\left(
 \begin{array}{cc}
 M_0- {\rm Re}(\delta') & M_{12}\\
 M_{12}^* & M_0+{\rm Re}(\delta')
 \end{array}
\right) - \frac{i}{2} \left(
 \begin{array}{cc}
 \Gamma_0+ 2{\rm Im}(\delta') & \Gamma_{12}\\
 \Gamma_{12}^* & \Gamma_0- 2{\rm Im}(\delta')
 \end{array}
\right)\right]\,,
\label{h-mat}
\ee
where $\delta'$ is defined by
\be
\delta = \frac{2\delta'}{\sqrt{H_{12}H_{21}}}\,.
\ee

One could even relax the assumption of $H_{21}=H_{12}^\ast$. However, there are two points that one
must note. First, the effect of expressing
$H_{12} = h_{12}+\bar\delta$, $H_{21} = h_{12}^\ast - \bar\delta$ appears as ${\bar\delta}^2$ in
$\sqrt{H_{12}H_{21}}$, the relevant expression in Eq.\ (\ref{deldef}), and can be neglected if we assume
$\bar\delta$ to be small. The second point, which is more important, is that CPT conservation constrains only
the diagonal elements and puts no constraint whatsoever on the off-diagonal elements.
It has been shown in \cite{lavoura} that $H_{12}\not= H_{21}^\ast$ leads to
T violation, and only $H_{11}\not= H_{22}$ leads to unambiguous CPT violation. Thus, we will
focus on the parametrization used in Eqs.\ (\ref{deldef}) and (\ref{h-mat}) to discuss the effects of CPT violation.

In the review on CPT violation in \cite{pdg}, the authors have used a formalism which is close to ours. While their
treatment is for the $K_S$-$K_L$ pair, this can be generalized to any neutral meson system. The mass
eigenstates are defined as
\be
|K_S (K_L)\ket = \frac{1}{\sqrt{2(1+|\epsilon_{s(L)}|^2)}} \left[ (1+\epsilon_{S(L)})|K^0\ket +
(1- \epsilon_{S(L)}) |\overline{K}{}^0\ket \right]
\ee
where
\bea
\epsilon_{S(L)} &=& \frac{ 
-i{\rm Im}(M_{12}) -\frac12{\rm Im}(\Gamma_{12}) \mp\frac12\left[M_{11}-M_{22} - \frac{i}{2}(\Gamma
_{11}-\Gamma_{22})\right]}
{M_L-M_S + i(\Gamma_S-\Gamma_L)/2} \nonumber\\
&\equiv& \epsilon \pm \tilde\delta\,.
\eea
Note that $\tilde\delta$ and $\delta$ are not the same, but related; 
both parametrize CPT violation. On the other hand, $\epsilon_{S(L)}$ is not truely a
CPT conserving quantity, as the expression contains the mass and width differences of the two eigenstates,
and both depend on the CPT violating parameter $\delta$ that we have used here.

The Belle collaboration \cite{cpt-belle} recently put stringent limits on the real and imaginary parts
of $\delta$,
\be
{\rm Re}(\delta_d) = (-3.8 \pm 9.9) \times 10^{-2}\,,\ \
{\rm Im}(\delta_d) = (1.14 \pm 0.93) \times 10^{-2}\,,
\ee
where we have added the errors in quadrature, and used the straightforward translation valid for
small $\delta$, {\em viz.}, $\delta = -2z$ (the subscript emphasizes that these results are for the $B_d$
system). The CPT violating parameter $z$ is defined as
\be
|B_{L(H)}\ket = p\sqrt{1-(+)z} |B^0\ket +(-) q\sqrt{1+(-)z}|\overline{B}{}^0\ket\,.
\ee
We can see that within the error bars data are consistent with no CPTV case i.e
${\rm Re}(\delta_d)={\rm Im}(\delta_d)=0$. However, more precise measurements are important and essential.
In any case it is safe to assume $|\delta|\ll 1$, even for the $B_s$ system.
In $\dms$ and $\dgs$ the CPT-violating effects are quadratic in $\delta$ and hence negligible.

%The eigenvalues of Eq. (\ref{h-mat}) are
%\bea
% \lambda &=& \left(M_0- \frac{i}{2}\Gamma_0\right) \pm H_{12}\alpha y
%\nonumber
%\\
%{\rm or},\ \   \lambda &=& \left[H_{11} + H_{12}\alpha \left(y +
%\frac{\delta}{2}\right)\right]\,,\ \
%%\nonumber\\
% \left[H_{22} - H_{12}\alpha \left(y + \frac{\delta}{2}\right)\right]\,,
%\eea
%where
%$y = \sqrt{ 1 + \frac{\delta^2}{4} }$ and
%$\alpha = \sqrt{ H_{21} / H_{12} }$.

We can write
\be
 |B_H\ket  = p_1|\bs\ket + q_1|\bsbar\ket\,,\ \
|B_L\ket = p_2|\bs\ket - q_2|\bsbar\ket\,.
\ee
with the normalization conditions
$|p_1|^2 + |q_1|^2 = |p_2|^2 + |q_2|^2 = 1$, so that with CPT violation,
$p_1\not= p_2$ and $q_1\not= q_2$.
The time evolutions of $B_H$ and $B_L$ are controlled by $\lambda_1\equiv
m_1-i\Gamma_1/2$ and $\lambda_2\equiv m_2-i\Gamma_2/2$ respectively. 
We also use
\be
\Delta M_s = m_1-m_2\,,\ \ \ 
\dgs = \Gamma_2-\Gamma_1\,.
\ee

Let us define,
\be
y = \sqrt{ 1 + \frac{\delta^2}{4} }\,;
\ \ \
 \eta_1 \equiv \frac{q_1}{p_1} = \left(y + \frac{\delta}{2}\right) \alpha\,;
\ \ \
\eta_2 \equiv \frac{q_2}{p_2} = \left(y - \frac{\delta}{2}\right) \alpha\,;
\ \ \
\omega = \frac{\eta_1}{\eta_2}\,,
\ee
where $\alpha= \sqrt{ H_{21} / H_{12} }$.
For $\vert\delta\vert \ll 1$, we can approximate $y$ with unity.

The time-dependent flavor eigenstates are given by
\bea
|\bs(t)\ket & = &h_+(t) |\bs\ket + \eta_1 h_-(t) |\bsbar\ket \nonumber\\
|\bsbar(t)\ket & =& \frac{h_-(t)}{\eta_2} |\bs\ket + \bar h_{+}(t)
|\bsbar\ket\,,
\eea
where
\bea
\nonumber h_-(t) & = &
\frac{1}{(1 + \omega)} \left(e^{-i\lambda_1 t} - e^{-i\lambda_2 t}\right)\,, \\
\nonumber h_+(t) & = &\frac{1}{(1 + \omega)} \left(e^{-i\lambda_1 t} + \omega e^{-i\lambda_2 t}\right)\,, \\
 \bar h_+(t) & = &\frac{1}{(1 + \omega)} \left(\omega e^{-i\lambda_1 t} + e^{-i\lambda_2 t}\right)\,.
\label{funct}
\eea
and we refer the reader to \cite{patra} for detailed expressions.
Note that in the absence of CPTV, $\eta_1=\eta_2$, $\omega=1$, and hence 
$h_+(t) = \bar h_+(t)$. In the limit $\vert\delta\vert \ll 1$, $\omega \approx 1+\delta$.

With our convention of  $|f\ket \equiv |D_s^+K^-\ket$ and $|\bar{f}\ket \equiv |D_s^- K^+\ket$,
where both the states are directly accessible to $\bs$ and $\bsbar$,
the time dependent decay rates are \cite{patra}
\bea
\Gamma(\bs(t)\rightarrow f) & = &
\left[|h_+(t)|^2 + |\xi_{f_1}|^2 |h_-(t)|^2 + 2{\rm Re}\left(\xi_{f_1} h_-(t) h_+^*(t)\right)\right] |A_f|^2\,, \nonumber\\
\Gamma(\bsbar (t)\rightarrow f) & = &
\left[|h_-(t)|^2 + |\xi_{f_2}|^2 |\bar h_+(t)|^2 + 2{\rm Re}\left(\xi_{f_2} \bar h_
+(t) h_-^*(t)\right)\right] \left|\frac{A_f}{\eta_2}\right|^2\,,\nonumber\\
%\eea
%
%{\bf{
%\bea
\Gamma(\bs(t)\rightarrow \bar f) & = &
\left[|h_+(t)|^2 + |\xi_{f_1}'|^2 |h_-(t)|^2 + 2{\rm Re}\left(\xi_{f_1}' h_-(t) h_+^*(t)\right)\right] |A_{\bar f}|^2\,, \nonumber\\
\Gamma(\bsbar (t)\rightarrow \bar f) & = &
\left[|h_-(t)|^2 + |\xi_{f_2}'|^2 |\bar h_+(t)|^2 + 2{\rm Re}\left(\xi_{f_2}' \bar h_+(t) h_-^*(t)\right)\right] 
\left|\frac{A_{\bar f}}{\eta_2}\right|^2\,,\label{taguntag}
\eea
where,
\be
\nonumber \xi_{f_1} = \eta_1 \frac{\bar A_f}{A_f} = \left(1 + \frac{\delta}{2}\right) \xi_f\,,\ \ \
\xi_{f_2} = \eta_2 \frac{\bar A_f}{A_f} = \left(1 - \frac{\delta}{2}\right) \xi_f\,,
\ee
\be
\xi_{f_1}' = \eta_1 \frac{\bar A_{\bar f}}{A_{\bar f}} = \left(1 + \frac{\delta}{2}\right) \xi_{\bar f}\,,\ \ \
\xi_{f_2}' = \eta_2 \frac{\bar A_{\bar f}}{A_{\bar f}} = \left(1 - \frac{\delta}{2}\right) \xi_{\bar f}\,.
\ee

Dropping terms ${\cal{O}}(\delta^2)$ or higher, we get the following expressions for
the tagged and untagged time-dependent decay rates:
\begin{align}
\nonumber  \Gamma(\bs(t)\rightarrow f) - \Gamma(\bsbar(t)\rightarrow f) & =
 \left[P_1 \sinh(\dgs t / 2) + Q_1 \cosh(\dgs t / 2) \right. \nonumber \\
& \left.  + R_1 \cos(\dms t) + S_1 \sin(\dms t)\right] e^{- \Gamma_s t} |A_f|^2\,, \nonumber\\
\Gamma(\bs(t)\rightarrow f) + \Gamma(\bsbar (t) \rightarrow f) & = \left[P_2 \sinh(\dgs t / 2) + Q_2
\cosh(\dgs t / 2) \right. \nonumber  \\
& \left. + R_2 \cos(\dms t) + S_2 \sin(\dms t)\right] e^{- \Gamma_s t} |A_f|^2\,,
\label{taguntagcpt}
\end{align}
with
\bea
P_1 &=& - \frac{1}{2}\, Re(\delta) \left(1 + |\xi_{f}|^2\right)\,,\nonumber\\
Q_1 &=&  - |\xi_{f}|\, \cos(\gamma + 2\beta_s +\Delta)\, {\rm Re}(\delta)\,, \nonumber\\
R_1 &=& 1-|\xi_{f}|^2 + |\xi_{f}|\, \cos(\gamma + 2\beta_s +\Delta) {\rm Re}(\delta)\,, \nonumber \\
S_1 &=& 2\,|\xi_{f}|\, \sin(\gamma + 2\beta_s +\Delta) - \frac{1}{2}\, {\rm Im}(\delta) \left(1 + |\xi_{f}|^2\right)\,,\nonumber\\
P_2 &=& 2\, |\xi_{f}|\, \cos(\gamma + 2\beta_s +\Delta)  - \frac{1}{2}\, {\rm Re}(\delta) \left(1 - 
|\xi_{f}|^2\right)\,,\nonumber\\
Q_2 &=& 1 + |\xi_f|^2 - |\xi_{f}|\, \sin(\gamma + 2\beta_s +\Delta) {\rm Im}(\delta)\,,\nonumber\\
 R_2 &=&  |\xi_{f}|\, \sin(\gamma + 2\beta_s +\Delta) Im(\delta)\,,\nonumber\\
S_2 &=& - \frac{1}{2}\, {\rm Im}(\delta) \left(1 - |\xi_{f}|^2\right)\,.
\label{obs1}
\eea
It is clear from Eq.\ (\ref{obs1}) that CPT violating effects in decay will not affect the determination of 
these 8 coefficients. Whatever the effects are, they will be lumped in the overall normalization $|A_f|^2$ and will
not appear in the coefficients of the trigonometric and hyperbolic functions.

All the 8 coefficients can theoretically be extracted from a fit to the time-dependent decay rates, but
admittedly the coefficients of the hyperbolic functions are harder to extract and need more statistics.  
The coefficients $P_1$ - $S_1$ are to be extracted from the tagged measurements, and $P_2$ - $S_2$ from
untagged measurements. 
Note that whether or not any CPT-conserving NP is present, absence of CPT violation definitely means $\delta=0$, so
$P_1 = Q_1 = R_2 = S_2 =0$. If any of these four observables are found to be nonzero, that is a sure signal
of CPT violation. (While $P_1$ and $S_2$ depend only on $\delta$, $Q_1$ and $R_2$ also have an implicit 
dependence on the $\bsbsbar$ mixing phase $2\beta_s$, which might depend on CPT conserving NP effects.) 
Therefore, if CPT is conserved, the tagged measurements are sensitive only to the 
trigonometric functions, and the untagged
measurements only to the hyperbolic functions, but we urge our experimental colleagues to perform a complete
fit.

If at least $P_1$ or $S_2$ be nonzero (maybe with nonzero $Q_1$ and $R_2$), one gets
\be
{\rm Im}(\delta)  = -\frac{2 S_2}{R_1 + Q_1}\,,  \hskip 30pt {\rm Re}(\delta)  = -\frac{2 P_1}{R_2 + Q_2},
\ee
which is theoretically clean, {\em i.e.} free from hadronic uncertainties. The overall normalization can be
extracted from the CP averaged branching fractions.

Even if the experiment is not sensitive enough to extract unambiguously nonzero values of
$P_1$, $Q_1$, $R_2$, or $S_2$, one can still find signals of CPTV, from the fact that
$P_2$, $Q_2$, $R_1$, and $S_1$ contain CPTV terms over and above CPT conserving but CP 
violating terms.
For example, one can extract the following analogous quantities from the tagged and
untagged $B_s\to\bar{f}$ decays:
\bea
&& \bar P_2 = 2\, |\xi_{f}|\,\cos(\gamma + 2\beta_s -\Delta)  + \frac{1}{2}\, {\rm Re}(\delta) \left(1 - 
|\xi_{f}|^2\right)\,,\nonumber \\
&& \bar Q_2 = 1 + |\xi_f|^2 - |\xi_{f}|\,\sin(\gamma + 2\beta_s -\Delta) {\rm Im}(\delta)\,,\nonumber\\
&& \bar R_1 = -1+|\xi_{f}|^2 + |\xi_{f}|\,\cos(\gamma + 2 \beta_s -\Delta) {\rm Re}(\delta)\,,\nonumber \\
&& \bar S_1 = 2\,|\xi_{f}|\,\sin(\gamma + 2 \beta_s -\Delta) - \frac{1}{2}\, {\rm Im}(\delta) \left(1 + |\xi_{f}|^2\right)\,.
\label{obs2}
\eea
%
%\hldc{If the initial strong phase assignment was wrong, every $-\Delta$ will change to $+\Delta$.}
It is easy to derive Eq.\ (\ref{obs2}) from Eq.\ (\ref{obs1}). First, note that the relevant expressions 
contain $|A_{\bar f}|$ and $\xi_{\bar f}$. Eq.\ (\ref{obs2}) follows when one substitutes $|\xi_
{\bar f}|=1/|\xi_f|$ and $|A_{\bar f}|^2/|\xi_f|^2 = |A_f|^2$. However, the strong phase changes sign
because of the definitions of $\xi_f$ and $\xi_{\bar f}$.

Therefore, from Eqs.~(\ref{obs1}) and (\ref{obs2}) we can define observables which are only sensitive to
the CPTV effect independent of any other NP effects in mixing,
\be
\frac{R_1 + {\bar R_1}}{P_2 + {\bar P_2}} = \frac{{\rm Re}(\delta)}{2}\,,
\hskip 20pt
\frac{Q_2 - {\bar Q_2}}{S_1 - {\bar S_1}} = \frac{{\rm Im}(\delta)}{2}\,.
\label{cptvp}
\ee
From Eq.~(\ref{cptvp}) we note that it is possible to probe the CPTV parameter $\delta$ even in the presence
of any other generic NP in mixing or decays (which modifies $2\beta_s$); the NP effects in mixing are cancelled in the ratio.
In addition we note that the strong phase is also exactly
cancelled in the ratio, hence the measurement of $\delta$ is free from hadronic uncertainties.

LHCb performs the decay profile fit assuming CPT invariance \cite{LHCb-029}, so it is not easy to 
predict the reach for the new CPT violating parameters, or even the CPT conserving ones. For this 
we need a full fit, assuming the possibility of CPT violation. Still, one can try to have an estimate
of the reach. As there exists no measurement on the CPT violating parameters, let us use the first 
relation of Eq.~(\ref{cptvp}). The parameter $R_1$ (called $C$ in \cite{LHCb-029}) has an error 
of about 56\% right now; if the data sample increases by a factor of 200, this might come down to 4\%. 
The same is true for $\bar{R_1}$, which should be measured independently (the central value, in the 
absence of CPT violation, should be equal and opposite to that of $R_1$). Thus the total uncertainty, 
added in quadrature, should be about 6\%. Similarly, the uncertainty in the denominator 
should be about 6\%, and is to be added in quadrature with the numerator. 
Thus, ${\rm Re}(\delta) \geq 0.16$ should be measurable using this relationship.
Of course, we expect a much better reach with a full 4-parameter fit to each decay profile. 
%  and this can be taken  if $\delta$ is small ($\sim
%0.1\%$ or smaller), we will possibly need a factor of 100 enhancement in the number of events to reach the
%sensitivity. This is not unrealistic; as we have discussed before, the full run will probably enhance the data set
%by a factor of about 200.

We reiterate that even if CPTV is present in decay, the conclusion that a nonzero value of
any one of the four observables $P_1$, $Q_1$, $R_2$, or $S_2$ indicates CPTV in mixing remains 
valid. Consider the expressions for the tagged and untagged decay rates, Eq.\ 
(\ref{taguntagcpt}). With enough satistics, one gets the coefficients of the trigonometric and the 
hyperbolic functions, as well as the overall normalization $\vert A_f\vert^2$.
If CPTV is present in decay, the expression for $\vert A_f\vert^2$ will change and be a function
of $y_f$, but the eight coefficients of Eq.\ (\ref{taguntagcpt}) will remain the same.

%While we have treated CPT violations in decay and mixing separately, the same method is applicable if
%CPTV is present in both. Consider the expressions for the tagged and untagged decay rates, Eq.\ 
%(\ref{taguntagcpt}). Similarly, 
%the expressions in Eqs. (\ref{amp01}) and (\ref{brfrac}) do not depend on these coefficients. 

The same method is applicable to decays like $B_s \to D_0 \phi$,  with
${\bar b} \to {\bar c} u {\bar s}$ and ${\bar b} \to {\bar u} c {\bar s}$ transitions.

%%%%%%%%%%%%%%%%%%%%%%%%%%%%%%%%%%%%%%%%%%%%%%%%%
\section{Summary and Conclusions}
%%%%%%%%%%%%%%%%%%%%%%%%%%%%%%%%%%%%%%%%%%%%%%%%%%%

While the effects of CPT violation are severely constrained for systems with first and/or second generation
fermions, the $B$ systems, in particular $B_s$, are relatively less constrained. This opens up the 
possibility of a CPT violating
action that is flavor-dependent. As a typical example of the effects of CPT violation, we consider the 
decays $\bs,\bsbar \to D_s^\pm K^\mp$. These decays are excellent probes of any NP; in the SM, they are
single-amplitude processes, and both $\bs\to D_s^+K^-$ and $\bs\to D_s^-K^+$ 
amplitudes are of the same order in Wolfenstein parametrization. 
Thus, the number of events for both $D_s^+K^-$ and $D_s^-K^+$, summing over parents $\bs$ and $\bsbar$,
should be the same in the SM. We show how this asymmetry becomes nonzero if there is CPT violation in the 
decay. 

At the same time, we see that if there is some NP that conserves CPT but comes with different strong
and weak phases from the corresponding SM amplitude, the asymmetry is again nonzero. So, while this asymmetry 
serves as an excellent indicator of any NP, it might be either CPT conserving (but necessarily CP 
violating) or CPT violating, and further
checks are necessary.

The situation is far better if there is CPT violation in mixing. The best way to put CPTV in mixing is to
make the diagonal terms of the $2\times 2$ mixing Hamiltonian unequal. With this, the CPTV parameter enters
the definition of the mass eigenstates, and through that, to various time-dependent decay rates. With
sufficient statistics, one can extract the coefficients of the trigonometric and hyperbolic terms of both
tagged and untagged time-dependent rates. We find that there are four coefficients which are zero not only in
the SM but also any extension with CPT conservation, so any nonzero value for any of them is a definite 
indication for CPT violation. There are several ways to extract these coefficients, and LHCb should have 
enough statistics to be able 
to measure them with sufficient precision. The argument goes through
even if CPTV is present in both decay and mixing; this is because different sets of observables are extracted for
the two different cases.

%While we have considered CPTV in either decay or mixing to simplify the formalism, the same argument 

\centerline{\bf{Acknowledgements}}

AK was supported by CSIR, Government of India, and also by
the DRS programme of the UGC, Government of India. The work of AS
was supported in part by the US DOE contract no.\ 
DE-AC02-98CH10886 (BNL). We acknowledge helpful discussions
with Tim Gershon (in particular about the LHCb reach), 
Vladimir Gligorov, and Tomasz Skwarnicki.

%%%%%%%%%%%%%%%%%%%%%%%%%%%

\end{document}